\documentclass[aps,pra,twocolumn,superscriptaddress,showpacs,nofootinbib,floatfix,10pt]{revtex4-1}

\usepackage[utf8]{inputenc}
\usepackage{microtype}

\usepackage{amsmath}
\usepackage{amssymb}
\usepackage{bm}

\renewcommand{\vec}{\bm}
\newcommand{\dif}{\mathrm{d}}
\newcommand{\me}{\mathrm{e}}
\newcommand{\mi}{\mathrm{i}}

\usepackage{graphicx}

\usepackage[pdfstartview=FitH,colorlinks=true,linkcolor=blue,citecolor=blue,urlcolor=blue]{hyperref}

\begin{document}

\title{Optical-lattice-assisted magnetic phase transition in a spin-orbit-coupled Bose-Einstein condensate}

\author{Giovanni I. Martone}
\email{Giovanni.Martone@ba.infn.it}
\affiliation{INO-CNR BEC Center and Dipartimento di Fisica, Universit\`{a} di Trento, I-38123 Povo, Italy}
\affiliation{Dipartimento di Fisica and MECENAS, Universit\`{a} di Bari, I-70126 Bari, Italy}
\affiliation{INFN, Sezione di Bari, I-70126 Bari, Italy}
\affiliation{LPTMS, CNRS, Univ. Paris-Sud, Universit\'{e} Paris-Saclay, 91405 Orsay, France}

\author{Tomoki Ozawa}
\affiliation{INO-CNR BEC Center and Dipartimento di Fisica, Universit\`{a} di Trento, I-38123 Povo, Italy}

\author{Chunlei Qu}
\affiliation{INO-CNR BEC Center and Dipartimento di Fisica, Universit\`{a} di Trento, I-38123 Povo, Italy}

\author{Sandro Stringari}
\affiliation{INO-CNR BEC Center and Dipartimento di Fisica, Universit\`{a} di Trento, I-38123 Povo, Italy}

\date{\today}

\begin{abstract}
We investigate the effect of a periodic potential generated by a one-dimensional optical lattice on the magnetic properties of an
$S=1/2$ spin-orbit-coupled Bose gas. By increasing the lattice strength one can achieve a magnetic phase transition between
a polarized and an unpolarized Bloch wave phase, characterized by a significant enhancement of the contrast of the density fringes.
If the wave vector of the periodic potential is chosen close to the roton momentum, the transition could take place at very small
lattice intensities, revealing the strong enhancement of the response of the system to a weak density perturbation. By solving
the Gross-Pitaevskii equation in the presence of a three-dimensional trapping potential, we shed light on the possibility of observing
the magnetic phase transition in currently available experimental conditions.
\end{abstract}


\maketitle

\section{Introduction}
\label{sec:introduction}
Spin-orbit-coupled Bose-Einstein condensates (BECs) are characterized by a rich variety of quantum phases,
which have already been the subject of theoretical and experimental investigations (see the recent
reviews~\cite{Dalibard2011review,Galitski2013review,Zhou2013review,Goldman2014review,Zhai2015review,
Li2015review,Zhang2016review} and references therein). In particular, in the case of a BEC of pseudospin $1/2$
with equal Rashba~\cite{Bychkov1984} and Dresselhaus~\cite{Dresselhaus1955} spin-orbit couplings, by tuning
the value of the Raman coupling between the two pseudospin states one can explore different phase transitions.
For relatively large values of the Raman coupling, of the order of the recoil energy, these systems exhibit a
second-order transition between a plane-wave and a single-minimum phase~\cite{Lin2011,Li2012PRL}.
The former is characterized by the macroscopic occupation of a single-particle state with finite momentum and
magnetic polarization, while in the latter the atoms populate an unpolarized state with vanishing momentum.
The above transition is associated with a divergent behavior of the magnetic susceptibility and with a large increase
of the effective mass. At a dynamic level it is characterized by the softening of the sound velocity~\cite{Martone2012,
Zheng2013,Ji2015} and of the frequency of the collective oscillations in the presence of harmonic trapping~\cite{Zhang2012,
Li2012EPL}. When one decreases the value of the Raman coupling the plane-wave phase eventually disappears in favor
of the so-called striped phase~\cite{Wang2010,Wu2011,Ho2011,Sinha2011,Li2012PRL,Ozawa2012,Li2013,Zezyulin2013,Lan2014,
Han2015,Natu2015,Sun2015,Liao2015,Sun2016,Yu2016,Martone2016}. In this configuration one has the appearance of periodic
modulations in the density profile, whose contrast depends on the value of the Raman coupling. In the absence of an
effective magnetic field the magnetic polarization vanishes in the striped phase. The phase transition between the
plane-wave and the striped phases has a first-order nature. It is characterized by the occurrence, on the side of the
plane-wave phase, of a roton minimum in the excitation spectrum, whose energy becomes smaller and smaller as one approaches
the transition~\cite{Martone2012,Zheng2013,Khamehchi2014,Ji2015}. The presence of the striped phase is one of the most
interesting features exhibited by spin-orbit-coupled BECs, due to its direct link to the long sought phenomenon of
supersolidity~\cite{Boninsegni2012}, which takes place when two continuous symmetries (gauge and translational invariance)
are spontaneously broken. So far the striped phase has not been identified directly in experiments because of the very small
contrast of the density fringes in the available experimental conditions~\cite{Martone2014}.

The occurrence of the roton excitation in the plane-wave phase and the emergence of the striped phase are deeply related
physical phenomena. This link becomes more evident by looking at the effects of a one-dimensional static periodic potential
applied to the BEC in the plane-wave phase. The linear response of the gas to a perturbation with wave vector close to the
roton momentum is greatly enhanced when approaching the transition to the striped phase~\cite{Martone2012}. Beyond the linear
regime, it has been shown that the application of the external potential can induce a magnetic phase transition to a fully
unpolarized configuration~\cite{Chen2016}. In this paper we investigate in detail the connection between the properties of
the response of the BEC at the linear and the nonlinear level. This results in a strong dependence of the behavior of the
system on the parameters of the external periodic potential. For example, when its wave vector is close to the roton minimum,
even a tiny static field is capable of inducing the transition to the unpolarized phase, with the appearance of highly contrasted
density modulations and strong magnetic fluctuations. For wave vectors different from the roton momentum, the magnetic phase
transition can take place at larger intensities of the optical lattice.

The paper is structured as follows. In Sec.~\ref{sec:soc_becs_unif} we review some relevant properties of the quantum phases
of a spin-orbit-coupled BEC in uniform matter and we point out the peculiar behavior of the static density response in the
plane-wave phase. Section~\ref{sec:soc_becs_ol} deals with the ground state of the system in the presence of a static optical
lattice; we mainly focus on the occurrence of a magnetic phase transition, which takes place at a critical lattice intensity whose
value depends in a nontrivial way on the Raman coupling and the lattice wave vector. The role of magnetic fluctuations and the effects
of an external trapping potential are also investigated. We summarize in Sec.~\ref{sec:conclusion}. Finally, in the Appendix we provide
a brief description of the features of the band structure of an ideal Bose gas with spin-orbit coupling.

\section{Spin-orbit-coupled Bose-Einstein condensates in uniform matter}
\label{sec:soc_becs_unif}
\subsection{Single-particle Hamiltonian}
\label{subsec:sp_ham}
The single-particle Hamiltonian first realized in the experiment of Ref.~\cite{Lin2011} reads (we set $\hbar = 1$)
\begin{equation}
h_0 = \frac{1}{2m} \left[ (p_x - k_0 \sigma_z)^2 + p_\perp^2 \right]
+ \frac{\Omega}{2}\sigma_x + \frac{\delta}{2}\sigma_z \, .
\label{eq:soc_ham}
\end{equation}
It acts on two-component spinors describing bosons with pseudospin up ($\uparrow$) and down ($\downarrow$);
for simplicity, we will refer to these degrees of freedom just as the two spin states of the system.
Equation~\eqref{eq:soc_ham} accounts for the presence of two counterpropagating and linearly polarized Raman lasers,
which provide transitions between different hyperfine levels of the atoms, and of a bias magnetic field.
The Raman coupling strength is quantified by $\Omega$, while $k_0$ is the momentum transfer due to the lasers.
The latter also fixes the value of the Raman recoil energy $E_r \equiv k_0^2 / 2m$. The linear Zeeman term $\delta$
represents an effective magnetic field, given by the sum of the Raman detuning and of the physical external magnetic
field (see, for example, Ref.~\cite{Martone2012}). Finally, $\sigma_i$ with $i=x,y,z$ denotes the $i$th Pauli matrix.
For vanishing $\delta$, when the Raman coupling $\Omega$ is smaller than $4E_r$, the lower branch of the single-particle
dispersion exhibits two degenerate minima at $p_x = \pm k_1^0$, with 
\begin{equation}
k_1^0 \equiv  k_0 \sqrt{1 - \left(\frac{\Omega}{4 E_r}\right)^2} \, ,
\label{eq:k10}
\end{equation}
while for $\Omega \geq 4 E_r$ it has one single minimum at $p_x = 0$.

\subsection{Many-body ground state}
\label{subsec:mb_ground_state}
We now discuss the effects of the two-body interactions. In the Gross-Pitaevskii mean-field approach, the energy
of an interacting system of $N$ spin-orbit-coupled bosons enclosed in a volume $V$ is given by
\begin{equation}
\begin{split}
E\left[\Psi\right] = \int_V \dif\vec{r} \big\{ & \Psi^\dagger(\vec{r}) h_0 \Psi(\vec{r}) \\
&{} + g_1 n^2(\vec{r}) + g_2 s_z^2(\vec{r}) + g_3 n(\vec{r}) s_z(\vec{r}) \big\} \, ,
\end{split}
\label{eq:soc_energy}
\end{equation}
where $\Psi$ denotes the two-component condensate wave function, $n(\vec{r}) = \Psi^\dagger(\vec{r}) \Psi(\vec{r})$
is the total density obeying the normalization condition $\int_V \dif\vec{r} \, n(\vec{r}) = N$, and $s_z(\vec{r}) =
\Psi^\dagger(\vec{r}) \sigma_z \Psi(\vec{r})$ is the longitudinal spin density. The coupling constants in
Eq.~\eqref{eq:soc_energy} correspond to the combinations $g_1 \equiv (g_{\uparrow\uparrow} + g_{\downarrow\downarrow}
+ 2 g_{\uparrow\downarrow})/8$, $g_2 \equiv (g_{\uparrow\uparrow} + g_{\downarrow\downarrow} - 2 g_{\uparrow\downarrow})/8$,
and $g_3 \equiv (g_{\uparrow\uparrow} - g_{\downarrow\downarrow})/4$ of the intraspecies and interspecies interaction strengths
$g_{\alpha\beta}$ ($\alpha,\beta = \uparrow,\downarrow$), which are related to the corresponding $s$-wave scattering lengths
$a_{\alpha\beta}$ via $g_{\alpha\beta} = 4\pi a_{\alpha\beta}/m$. In the following, unless otherwise specified, we will assume
equal intraspecies interactions $g_{\uparrow \uparrow} = g_{\downarrow \downarrow}$, yielding $g_3 = 0$. We will also take
$\delta = 0$, although the effects of a nonvanishing effective magnetic field will also be discussed qualitatively.

In the presence of antiferromagnetic spin-dependent interactions ($g_2 > 0$), when the Raman coupling $\Omega$ is small,
the ground state of the many-body system corresponds to the so-called striped phase~\cite{Wang2010,Wu2011,Ho2011,Sinha2011,
Li2012PRL,Ozawa2012,Li2013,Zezyulin2013,Lan2014,Han2015,Natu2015,Sun2015,Liao2015,Sun2016,Yu2016,Martone2016}. In this
configuration the density profile of the gas exhibits periodic modulations in the form of stripes, which appear as a consequence
of the spontaneous breaking of translational invariance. Another relevant feature of the striped phase is the absence of magnetic
polarization along $z$, that is, $\langle \sigma_z \rangle \equiv \int_V \dif\vec{r} \, s_z(\vec{r}) = 0$.

By increasing $\Omega$ the system enters the polarized plane-wave phase. In this phase the condensate order parameter is given
by the plane-wave function $\exp(\mi k_1 x)$ [or $\exp(- \mi k_1 x)$] times a real spinor. The components of the latter have
relative weights fixed by the value of the Raman coupling. The plane-wave phase is characterized by a uniform density and by a
finite value $\langle \sigma_z \rangle = N k_1/k_0$ (or $\langle \sigma_z \rangle = - N k_1/k_0$) of the longitudinal magnetic
polarization~\cite{Ho2011,Li2012PRL}. The momentum $k_1$ in the previous formulas does not actually coincide with its
single-particle value $k_1^0$ given by Eq.~\eqref{eq:k10} because of the spin-dependent interactions proportional to $g_2$;
one finds~\cite{Li2012PRL}
\begin{equation}
k_1 = k_0 \sqrt{1 - \left[\frac{\Omega}{4(E_r - \bar{n}g_2)}\right]^2} \, ,
\label{eq:k1}
\end{equation}
with $\bar{n} \equiv N/V$ the average density of the gas. To simplify the discussion, in this work we will not account for the small
difference between $k_1$ and $k_1^0$, which is negligible in the conditions of current experiments. The twofold degeneracy of the
plane-wave phase highlighted above stems from a spontaneous breaking mechanism of two $\mathbb{Z}_2$ symmetries of the energy
functional~\eqref{eq:soc_energy}, similarly to what happens in usual ferromagnetic configurations. More specifically, if both $\delta$
and $g_3$ are vanishing, the energy~\eqref{eq:soc_energy} is invariant under the separate action of $\sigma_x \mathcal{P}$
and $\sigma_z \mathcal{T}$, with $\mathcal{P}$ and $\mathcal{T}$ the parity and time-reversal operator, respectively. The plane-wave
phase breaks both the above symmetries, while being invariant under their product
$(\sigma_x \mathcal{P})(\sigma_z \mathcal{T})$.\footnote{Notice that the energy functional~\eqref{eq:soc_energy} keeps its invariance
under $(\sigma_x \mathcal{P}) (\sigma_z \mathcal{T})$ even if $\delta$ and $g_3$ are nonvanishing.}

The transition between the striped and the plane-wave phases occurs at a critical Raman coupling $\Omega^0_\mathrm{tr}$ whose value is given,
in the low-density limit, by the density-independent expression~\cite{Ho2011,Li2012PRL}
\begin{equation}
\Omega^0_\mathrm{tr} = 4E_r \sqrt{\frac{2 g_2}{g_1+2 g_2}} \, .
\label{eq:Omega_tr_0}  
\end{equation}
This transition is of first order and has an important magnetic character~\cite{Martone2012,Li2012EPL} that has been pointed
out experimentally in~\cite{Lin2011,Ji2014}. At even larger values of the Raman coupling $\Omega$ the BEC undergoes a
second-order transition to the single-minimum phase, in which the condensation momentum and the magnetic polarization are
both vanishing.\footnote{This description of the phase diagram of a spin-orbit-coupled BEC with $g_2 > 0$ holds for densities
$\bar{n}$ smaller than the critical value $\bar{n}_\mathrm{cr} = E_r g_1/[2 g_2 (g_1 + g_2)]$. For $\bar{n} > \bar{n}_\mathrm{cr}$
the plane-wave phase disappears and one has a direct first-order transition between the striped and the single-minimum phases.}
We finally mention that, if the spin-dependent interactions have a ferromagnetic nature ($g_2 < 0$), only the plane-wave
and the single-minimum phases appear in the phase diagram of the BEC, the striped phase being always energetically unfavored.

\subsection{Rotonic excitations and compressibility}
\label{subsec:roton_dens_resp}
\begin{figure}
\includegraphics{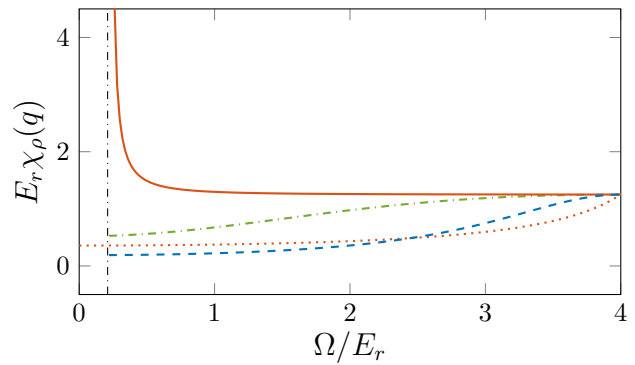}
\caption{(Color online) Static density response $\chi_\rho(q)$ of a spin-orbit-coupled BEC in the plane-wave phase as a function of
the Raman coupling $\Omega$, at three different values $q = 1.5k_1^0$ (green dash-dotted line), $q = 2k_1^0$ (red solid line),
and $q = 3k_1^0$ (blue dashed line) of the momentum transfer. The red dotted curve corresponds to the response at $q = 2k_1^0$
of a standard Bose gas without spin-orbit coupling. The vertical black dash-dotted line identifies the critical Raman coupling
$\Omega^0_\mathrm{tr}$ at which the transition to the striped phase takes place. The other parameters are $\bar{n} g_1 / E_r = 0.4$
and $g_2/g_1 = 0.0012$.}
\label{fig:chi_dens_x_q}
\end{figure}

A peculiar property of the plane-wave phase is the existence of a low-energy roton minimum in the excitation spectrum, occurring
at a momentum close to $2k_1$~\cite{Martone2012,Zheng2013,Khamehchi2014,Ji2015}. The energy of the roton becomes smaller and smaller
as $\Omega$ approaches (from above) the critical value $\Omega^0_\mathrm{tr}$, providing the onset of the transition to the striped
phase. The roton minimum is responsible for a peculiar behavior of the static density response, i.e., the compressibility $\chi_\rho(q)$.
According to the linear response theory, the compressibility can be calculated by adding, to the single-particle
Hamiltonian~\eqref{eq:soc_ham}, a small static periodic perturbation of the form $V_\lambda = -\lambda \rho_q + \mathrm{H.c.}$,
where $\rho_q \equiv \int_V \dif \vec{r} \, \me^{-\mi q x} \hat{n}(\vec{r})$ and $\hat{n}(\vec{r})$ is the density operator. Then
$\chi_\rho(q) \equiv \lim_{\lambda \to 0} \langle \rho_q \rangle_\lambda/\lambda$, where $\langle \rho_q \rangle_\lambda$ is the expectation
value of $\rho_q$ on the perturbed ground state. One finds that the function $\chi_\rho(q)$ exhibits a significant enhancement
when $q$ is close to the roton minimum~\cite{Martone2012}. A similar effect is known to characterize the static response of superfluid
helium~\cite{Dalfovo1992}. The enhancement becomes particularly strong when one approaches the phase transition to the striped phase at $\Omega =
\Omega^0_\mathrm{tr}$. In Fig.~\ref{fig:chi_dens_x_q} we plot the compressibility calculated at $q = 2k_1^0$ as a function of $\Omega$,
in the presence (red solid line) and in the absence (red dotted line) of spin-orbit coupling. The values of the average density $\bar{n}$
and of the ratio $g_2/g_1 = 0.0012$ employed in the figure correspond to current experiments with $^{87}$Rb BECs (see also the discussion
in Sec.~\ref{subsec:harm_trap}). In such conditions, the low-density expression~\eqref{eq:Omega_tr_0} yields the value
$\Omega^0_\mathrm{tr}/E_r = 0.19$ for the critical Raman coupling, which is raised to $\Omega^0_\mathrm{tr}/E_r = 0.21$ after including
the small corrections due to the finite density of the system. One can notice that, for small $\Omega$, the static response of a
spin-orbit-coupled BEC at $q = 2k_1^0$ takes extremely high values as compared to those of a Bose gas without spin-orbit coupling.
In the latter case the value of $\chi_\rho(q=2k_1^0)$ for small $\Omega$ is well approximated by the asymptotic large-$q$ behavior
$\chi_\rho(q) \to 4m/q^2$. The strong enhancement of $\chi_\rho(q)$ does not take place if the momentum transfer $q$ is significantly
smaller or larger than the roton momentum; to show this, in Fig.~\ref{fig:chi_dens_x_q} we also report the values of the compressibility
in the presence of spin-orbit coupling at $q = 1.5k_1^0$ (green dash-dotted line) and $q = 3k_1^0$ (blue dashed line).

The significant dependence of the static response on the momentum transfer is reflected in the peculiar behavior of the system
when entering the nonlinear regime. In particular, the large value of the compressibility close to the roton momentum suggests
that nonlinear effects will emerge soon in the response, even in the presence of a tiny periodic perturbation. They are expected
to give rise to highly contrasted density modulations and to important effects in the magnetic polarization of the gas, as we will
discuss in the next section.

\section{Magnetic phase transition in the presence of an optical lattice}
\label{sec:soc_becs_ol}

\subsection{Variational ansatz for the ground state}
\label{subsec:var_ansatz}
Let us now explore the behavior of a spin-orbit-coupled BEC when a one-dimensional sinusoidal periodic potential of the form
\begin{equation}
V(\vec{r}) = s E_\mathrm{latt} \sin^2 \frac{q x}{2}
\label{eq:periodic_pot}
\end{equation}
is added to the single-particle Hamiltonian~\eqref{eq:soc_ham}. In the previous expression, $q$ denotes the momentum transferred
by the two lasers, oriented along the $x$ axis, which generate the optical lattice. The dimensionless parameter $s$ fixes the
strength of the periodic potential in units of the lattice recoil energy $E_\mathrm{latt} \equiv (q/2)^2/2m$. We will focus on
the case of relatively weak lattice intensities, such that the Gross-Pitaevskii mean-field description is still applicable.

There have been several studies on spin-orbit-coupled BECs in shallow optical lattices~\cite{Larson2010,Sakaguchi2013,Zhang2013,
Kartashov2013,Cheng2014,Han2015,Salerno2015,Hamner2015,Li2015,Zhang2015,Poon2015,Chen2016,Hurst2016}. In particular,
the ground state in the presence of the periodic potential~\eqref{eq:periodic_pot} has been investigated in Ref.~\cite{Chen2016}.
In this work we focus on the properties of a magnetic phase transition induced by the periodic potential.

First, we recall that, if the atoms condense in a state with a well-defined value of the quasimomentum $k$, the condensate
wave function can be written in the usual Bloch form
\begin{equation}
\tilde{\Psi}_k\left(\vec{r}\right)
= \sqrt{\bar{n}} \, \me^{\mi k x} \sum_{\bar{K}} C_{k+\bar{K}}
\begin{pmatrix}
\phantom{+}\cos\theta_{k+\bar{K}} \\
- \sin\theta_{k+\bar{K}}
\end{pmatrix}
\me^{\mi \bar{K} x}.
\label{eq:bloch_waves}
\end{equation}
Here $\bar{K}$ are the reciprocal lattice vectors having values $\left\{\bar{m} q\right\}_{\bar{m}\in\mathbb{Z}}$,
while $\theta_{k+\bar{K}}$ and $C_{k+\bar{K}}$ are parameters characterizing the coefficients of the Bloch wave expansion.
The coefficients $C_{k+\bar{K}}$ satisfy the constraint $\sum_{\bar{K}} | C_{k+\bar{K}} |^2 = 1$, which ensures that
the order parameter~\eqref{eq:bloch_waves} is normalized to the total number of particles in the condensate, i.e.,
$\int_V \dif\vec{r} \, \tilde{\Psi}_k^\dagger(\vec{r}) \tilde{\Psi}_k(\vec{r}) = N$. Henceforth we will take $k$ in the
first Brillouin zone.

A good starting point for the study of the ground state of the BEC is represented by the properties of the band structure
of the system in the noninteracting limit~\cite{Zhang2013,Chen2016}, which are summarized in the Appendix. In particular,
in the first Brillouin zone the lowest-lying energy band exhibits two degenerate minima, which occur at two opposite finite values
of the quasimomentum [see Figs.~\ref{fig:band_structure}(a) and \ref{fig:band_structure}(b)]. These considerations suggest
that one can write an ansatz for the ground-state wave function as a superposition of two Bloch waves of the form~\cite{Chen2016} 
\begin{equation}
\Psi\left(\vec{r}\right) =
\tilde{C}_+ \tilde{\Psi}_{+k_s}\left(\vec{r}\right)
+ \tilde{C}_- \tilde{\Psi}_{-k_s}\left(\vec{r}\right) \, .
\label{eq:ansatz}
\end{equation}
Here $|\tilde{C}_+|^2 + |\tilde{C}_-|^2 = 1$ and the expressions of $\tilde{\Psi}_{\pm k_s}$ are given by
Eq.~\eqref{eq:bloch_waves} with $k = \pm k_s$, $k_s$ being the magnitude of the quasimomentum in the ground state.
For an ideal Bose gas, Eq.~\eqref{eq:ansatz} is able to reproduce the exact ground-state wave function of the
system; in particular, the quasimomenta $\pm k_s$ correspond to the minima of the lowest band mentioned above, while
the relative values of the coefficients $\tilde{C}_+$ and $\tilde{C}_-$ remain arbitrary.

The same ansatz~\eqref{eq:ansatz} is also well suited to study the ground state of the system in the presence of two-body
interactions~\cite{Chen2016}. Before proceeding, it is worth noticing that it allows one to recover all the results presented
in Sec.~\ref{sec:soc_becs_unif} for the quantum phases of the BEC in the absence of the optical lattice (i.e., for $s=0$).
In this limit the Bloch wave~\eqref{eq:bloch_waves} has $C_k = 1$ and $C_{k+\bar{K}} = 0$ for $\bar{K} \neq 0$, that is,
it reduces to a simple plane wave with momentum $k$. Consequently, the wave function~\eqref{eq:ansatz} becomes a
superposition of two counterpropagating plane waves with momenta $\pm k_s$, which coincides with the ansatz
employed in the variational analysis of Ref.~\cite{Li2012PRL}. The plane-wave phase is reproduced by setting
$k_s = k_1$, with $k_1$ given by Eq.~\eqref{eq:k1}, $\tilde{C}_+ = 1$, and $\tilde{C}_- = 0$ (or $\tilde{C}_+ = 0$
and $\tilde{C}_- = 1$ for the degenerate state with opposite momentum and magnetic polarization); if additionally $k_1$
vanishes, one gets instead the single-minimum phase. The striped phase corresponds to the choice $|\tilde{C}_+| = |\tilde{C}_-|
= 1/\sqrt{2}$.

When the optical lattice is turned on one expects that also the plane-wave components with $\bar{K} \neq 0$ in Eq.~\eqref{eq:bloch_waves}
are populated. In order to study the ground state of the system at finite $s$, we first insert the ansatz~\eqref{eq:ansatz} into the energy
functional~\eqref{eq:soc_energy} and we deduce an expression for the energy as a function of the variational parameters $k_s$,
$\theta_{\pm k_s+\bar{K}}$, $C_{\pm k_s+\bar{K}}$, and $\tilde{C}_\pm$. Then we minimize the energy at fixed values of $k_0$,
$\Omega$, $\bar{n}$, the ratio $g_2/g_1$ of the interaction strengths, and the lattice parameters $q$ and $s$. We have checked that the
results given by this variational procedure agree with those obtained by directly solving the Gross-Pitaevskii equation in a box configuration.

\subsection{Mixed regime}
\label{subsec:mixed_regime}
At sufficiently small values of the Raman coupling $\Omega$, if $g_2 > 0$, the ground-state wave function of the BEC contains both the Bloch wave
terms in the superposition~\eqref{eq:ansatz} with equal weights $|\tilde{C}_+| = |\tilde{C}_-| = 1/\sqrt{2}$, giving rise to the so-called mixed
phase. The properties of this phase have been studied in detail in Ref.~\cite{Chen2016}. Here we mention that it is characterized by vanishing magnetic
polarization and by modulations in the density profile at two different wavelengths: The first one, equal to $2\pi/q$, is fixed by the external lattice
potential, while the second one arises because of the spin-orbit coupling and is given by $\pi/k_s$. The presence of the latter entails a spontaneous
breaking of the discrete translational symmetry exhibited by the energy functional~\eqref{eq:soc_energy} after the addition of the lattice
potential~\eqref{eq:periodic_pot}. Depending on whether the two wavelengths are commensurate or not, the global oscillation of the density can be periodic
or nonperiodic. The ansatz~\eqref{eq:ansatz} actually provides only a first approximation for the wave function of the mixed phase, as it neglects
the higher-order Bloch wave terms caused by the nonlinear interactions in the energy functional~\eqref{eq:soc_energy}~\cite{Li2013,Chen2016}.
Notice that at $s = 0$ the mixed phase smoothly connects to the striped phase, where only modulations with period $\pi/k_s$ are present.

\subsection{Unmixed regime: Magnetic phase transition}
\label{subsec:unmixed_regime}
The situation becomes dramatically different if one increases $\Omega$. Above a critical value $\Omega_\mathrm{tr}$ of the Raman coupling, the mixed
phase becomes energetically unfavored and a first-order transition occurs to an unmixed regime. In the latter the procedure of energy minimization yields,
for the coefficients in the ansatz~\eqref{eq:ansatz}, the values $\tilde{C}_+ = 1$ and $\tilde{C}_- = 0$ or the values $\tilde{C}_+ = 0$ and $\tilde{C}_- = 1$,
the two choices corresponding to the same value of the energy~\cite{Chen2016}. Hence, in the unmixed regime the wave function of the BEC is given by
a single Bloch wave of the kind~\eqref{eq:bloch_waves}, whose quasimomentum can be assumed equal to either $+k_s$ or $-k_s$. The value of $\Omega_\mathrm{tr}$
separating the mixed and unmixed configurations depends on the lattice strength $s$ (see Fig.~\ref{fig:phase_diag_Omega_s}); for $s = 0$, it reduces
to the critical Raman coupling $\Omega^0_\mathrm{tr}$ at which the transition from the striped to the plane-wave phase occurs [see Eq.~\eqref{eq:Omega_tr_0}
and the related discussion]. In the rest of the present paper we will focus principally on the physics of the unmixed regime, which represents
the central part of our work.

The quantum phases appearing in the unmixed regime, which we discuss below, display modulations in the density profile with the same periodicity
$2\pi/q$ as the external lattice potential~\eqref{eq:periodic_pot}. For such periodic fringes one can define the contrast as
\begin{equation}
I \equiv \frac{n_\mathrm{max} - n_\mathrm{min}}{n_\mathrm{max} + n_\mathrm{min}} \, ,
\label{eq:contrast}
\end{equation}
where $n_\mathrm{max}$ and $n_\mathrm{min}$ are the maximum and the minimum value, respectively, taken by the density during its spatial oscillations.

\begin{figure}
\includegraphics{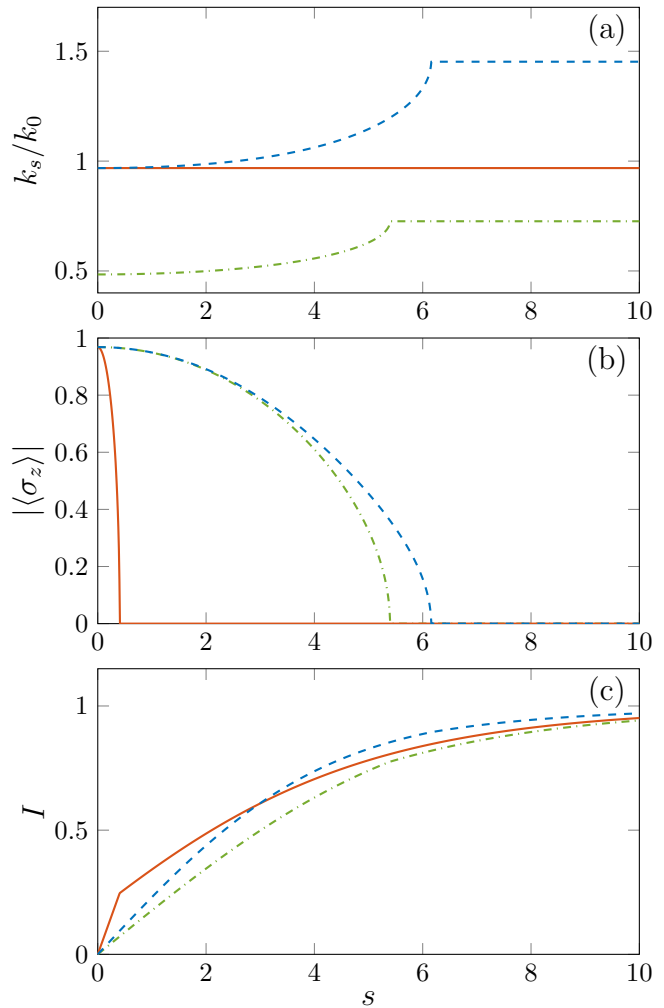}
\caption{(a) Condensation quasimomentum, (b) magnetic polarization, and (c) contrast of the fringes as a function of the dimensionless lattice
strength $s$ for $\Omega/E_r = 1.0$. In each panel we show the results for three different values $q = 1.5k_1^0$ (green dash-dotted line),
$q = 2k_1^0$ (red solid line), and $q = 3k_1^0$ (blue dashed line) of the momentum transfer. The other parameters are $\bar{n} g_1 / E_r = 0.4$
and $g_2/g_1 = 0.0012$.}
\label{fig:ks_sz_contr_bulk}
\end{figure}

For a fixed value of the Raman coupling $\Omega > \Omega_\mathrm{tr}$, the properties of the ground state are determined by the competition
between the density-density interaction term in the energy~\eqref{eq:soc_energy}, proportional to $g_1$, and the lattice potential with strength $s$. 
For very small values of $s$, the interactions favor a configuration where the atoms dominantly occupy the $\bar{K} = 0$ state in the
superposition~\eqref{eq:bloch_waves}, the populations $|C_{\pm k_s+\bar{K}}|^2$ of the terms with $\bar{K} \neq 0$ being much smaller. In this phase,
which smoothly connects to the plane-wave phase at $s = 0$, the ground state is twofold degenerate. In particular, the magnetic polarization $\left\langle \sigma_z
\right\rangle$ is finite and takes opposite values in the two states with quasimomentum $+k_s$ and $-k_s$. As in the plane-wave phase at $s = 0$, this
degeneracy stems from the spontaneous breaking of the $\sigma_x \mathcal{P}$ and $\sigma_z \mathcal{T}$ symmetries discussed in Sec.~\ref{sec:soc_becs_unif},
which are not affected by the addition of the periodic potential~\eqref{eq:periodic_pot}. Notice that the quasimomentum $k_s$ approaches $k_1$
[see Eq.~\eqref{eq:k1}] in the limit of vanishing lattice strength, provided that $k_1$ belongs to the first Brillouin zone, i.e., $k_1 \leq q/2$; more in general,
if the value of $k_1$ falls in the $\ell$th Brillouin zone, then $k_s$ tends to $|k_1 - (\ell - 1) q|$ [see, for example, the results for $q = 1.5k_1^0$ in
Fig.~\ref{fig:ks_sz_contr_bulk}(a), for which $\ell = 2$]. We will refer to the state described above as the polarized Bloch-wave phase.

As one increases $s$, the states with $\bar{K} \neq 0$ in Eq.~\eqref{eq:bloch_waves} become more and more populated at the expense
of the $\bar{K} = 0$ state. At the same time, the quasimomentum $k_s$, where Bose-Einstein condensation takes place, moves in
the direction of the wave vector $q/2$, dictated by the optical potential~\cite{Zhang2013,Chen2016}.\footnote{The last statement is true
as long as the value of $q$ is not much smaller or much larger than $2k_1^0$. In the opposite case a different behavior can take place, i.e.,
the condensation quasimomentum $k_s$ moves toward the center of the Brillouin zone at $k=0$. This effect never occurs for the values
of $q$ considered in the present work, hence we will not discuss it further.} This results in the decrease of the magnitude of the magnetic
polarization $\langle\sigma_z\rangle$, as well as in the increase of the contrast of the fringes~\eqref{eq:contrast}. In Fig.~\ref{fig:ks_sz_contr_bulk}
we plot these quantities as a function of $s$, for a fixed value $\Omega/E_r = 1.0$ of the Raman coupling and for several choices of the lattice wave
vector. In particular, for $q = 2k_1^0$ (red solid line), i.e., for a momentum transfer very close to the roton wave vector, the decrease of the magnetic
polarization occurs very rapidly as one increases the optical lattice strength $s$. This behavior of the nonlinear response of the system to the external
lattice potential~\eqref{eq:periodic_pot} is deeply connected to the enhancement of the compressibility discussed in Sec.~\ref{subsec:roton_dens_resp}.
Notice also that the quasimomentum $k_s$ lies extremely close to the edge of the first Brillouin zone for any $s$ in the $q = 2k_1^0$ case. Instead, for
$q = 1.5k_1^0$ (green dash-dotted line) and $q = 3k_1^0$ (blue dashed line) the reduction of $|\langle\sigma_z\rangle|$ and the shift of $k_s$ towards
the value $q/2$ take place for larger lattice intensities.

At the critical lattice strength $s_\mathrm{cr}$, whose value depends on $\Omega$ (see Fig.~\ref{fig:phase_diag_Omega_s}),
the condensation quasimomentum $k_s$ coincides with the wave vector of the periodic potential $q/2$, i.e., with the edge of
the first Brillouin zone, as shown in Fig.~\ref{fig:ks_sz_contr_bulk}(a). Since the two Bloch states having quasimomentum $\pm q/2$
are physically identical, the corresponding magnetic polarization must vanish [see Fig.~\ref{fig:ks_sz_contr_bulk}(b)]. As a consequence,
the system undergoes a second-order transition to an unpolarized Bloch wave phase. In this latter phase the populations of the various
plane-wave states of Eq.~\eqref{eq:bloch_waves} are balanced such that $|C_{-q/2-\bar{K}}|^2 = |C_{q/2+\bar{K}}|^2$ and
$\theta_{-q/2-\bar{K}} = \pi/2 - \theta_{q/2+\bar{K}}$ for any $\bar{K}$; hence, the $\sigma_x \mathcal{P}$ and the
$\sigma_z \mathcal{T}$ symmetries are restored in the unpolarized state. Concerning the contrast of the density
modulations~\eqref{eq:contrast}, which we plot in Fig.~\ref{fig:ks_sz_contr_bulk}(c), it keeps growing with $s$, although more slowly
than in the polarized phase. Notice that in the unpolarized Bloch wave phase the contrast is much larger than the typical values exhibited
in the striped phase in the absence of the lattice~\cite{Martone2014}; this is true even in the case $q = 2k_1^0$, where the magnetic phase
transition, for small values of $\Omega$ [see Fig.~\ref{fig:phase_diag_Omega_s}(b)], takes place at extremely small values $s$ of the optical
lattice strength.

\subsection{Phase diagram}
\label{subsec:phase_diag}
\begin{figure}
\includegraphics{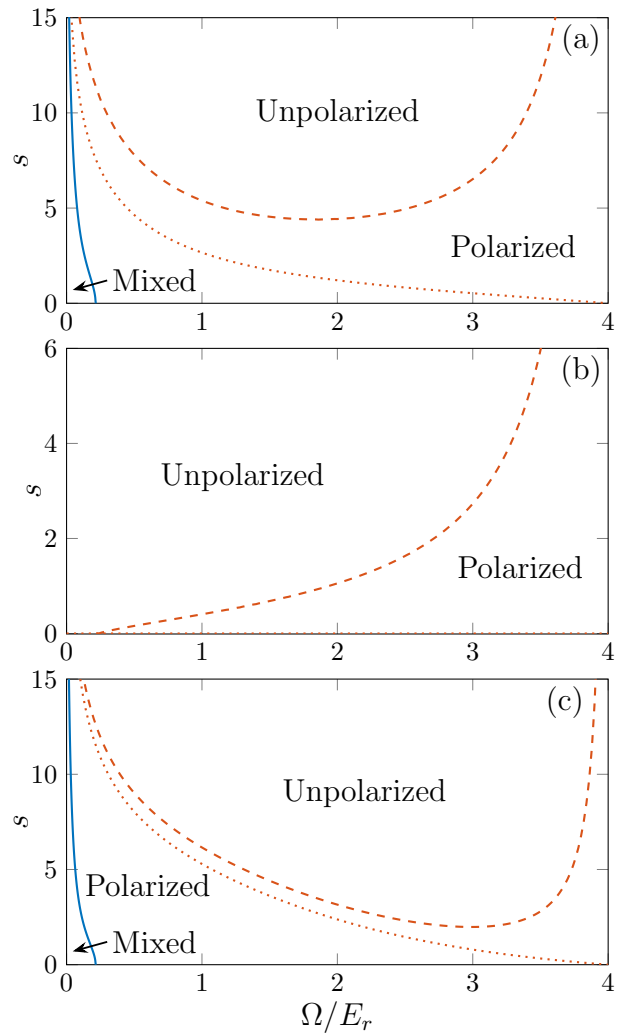}
\caption{(Color online) Phase diagram as a function of the Raman coupling $\Omega$ and of the dimensionless lattice strength $s$
for (a) $q = 1.5k_1^0$, (b) $q = 2k_1^0$, and (c) $q = 3k_1^0$. The blue solid line represents the first-order transition from
the mixed to the polarized Bloch wave phase. The red dashed line corresponds to the second-order transition from the polarized
to the unpolarized Bloch wave phase. For the latter, the prediction of the ideal gas model is also given (red dashed line).
The other parameters are $\bar{n} g_1/ E_r = 0.4$ and $g_2/g_1 = 0.0012$.}
\label{fig:phase_diag_Omega_s}
\end{figure}

In Fig.~\ref{fig:phase_diag_Omega_s} we show the phase diagram of the system in the $\Omega$-$s$ plane, for the same values of the average
density $\bar{n}$ and of the interaction parameters $g_1$ and $g_2$ as the previous figures. Each panel of Fig.~\ref{fig:phase_diag_Omega_s}
corresponds to a different value of $q$ [$q = 1.5k_1^0$ in Fig.~\ref{fig:phase_diag_Omega_s}(a), $q = 2k_1^0$
in Fig.~\ref{fig:phase_diag_Omega_s}(b), and $q = 3k_1^0$ in Fig.~\ref{fig:phase_diag_Omega_s}(c)]. In each diagram, the blue solid line
indicates the critical Raman coupling $\Omega_\mathrm{tr}$ at which, for a given value of $s$, the first-order transition from the mixed to
the polarized Bloch wave phase occurs. The red dashed lines show instead the behavior, as a function of $\Omega$, of the critical lattice
strength $s_\mathrm{cr}$ separating the polarized and unpolarized Bloch wave states. The predictions for $s_\mathrm{cr}$ obtained using the
ideal Bose gas model (see the Appendix) are also reported (red dotted lines).

The mixed phase discussed in Sec.~\ref{subsec:mixed_regime} appears in a very narrow region close to the left edge of the diagrams of
Figs.~\ref{fig:phase_diag_Omega_s}(a) and \ref{fig:phase_diag_Omega_s}(c). The value of $\Omega_\mathrm{tr}$ is maximum at $s = 0$,
where it coincides with $\Omega^0_\mathrm{tr}$, and becomes smaller and smaller as the lattice intensity $s$ increases. 

The rest of the diagram is occupied by the unmixed polarized and unpolarized configurations. One can notice that, if interactions are included,
the critical lattice strength $s_\mathrm{cr}$ shows a nonmonotonic behavior as a function of $\Omega$ in the $q = 1.5k_1^0$ and $q = 3k_1^0$
cases. The large values exhibited by $s_\mathrm{cr}$ at small $\Omega$ can be understood by recalling that, in this limit, the system is strongly
polarized in the absence of the lattice; hence, large lattice intensities are required to achieve the transition to the unpolarized phase.
In this regime of small Raman coupling the contribution of the interaction energy, being of the order of $\bar{n} g_1$ [see Eq.~\eqref{eq:soc_energy}],
is less important than that of the lattice potential~\eqref{eq:periodic_pot}, proportional to $s E_\mathrm{latt}$; consequently, the interactions
do not change qualitatively the behavior with respect to the prediction of the ideal Bose gas model (dotted lines). In the opposite limit of large
$\Omega$, the transition takes place at low values of $s$ in the ideal gas model. Since $k_1^0/k_0 \ll 1$, the energy scale $E_\mathrm{latt}$
associated with the lattice potential is very small and the interactions play an important role in determining the properties of the gas.
This leads to an increase of $s_\mathrm{cr}$ with respect to the values obtained within the ideal Bose gas model.

As $q \to 2 k_1^0$, from both above and below, the regions of the mixed and of the unpolarized Bloch wave phases get closer and closer and eventually
they touch, without the occurrence of the polarized Bloch wave phase in between. This process results in a different behavior of the phase diagram of
Fig.~\ref{fig:phase_diag_Omega_s}(b) for $q = 2k_1^0$ with respect to those of Figs.~\ref{fig:phase_diag_Omega_s}(a) and \ref{fig:phase_diag_Omega_s}(c).
In particular, the value of $s_\mathrm{cr}$ exhibits a monotonic increasing dependence on $\Omega$ for $\Omega > \Omega^0_\mathrm{tr}$. The difference
with respect to the $q = 1.5k_1^0$ and $q = 3k_1^0$ cases can be better understood by taking into account that the ideal gas model, for $q = 2k_1^0$,
gives rise to an unpolarized phase as soon as $s \neq 0$. The increase of $s_\mathrm{cr}$ with $\Omega$ is hence a pure effect of the two-body
interactions. In Fig.~\ref{fig:phase_diag_Omega_s}(b) we have not included the line separating the mixed and the unpolarized Bloch wave phases; its
calculation would require an accurate estimate of the energy difference between such phases, which, however, is extremely small if $q = 2k_1^0$ [this is
due to the closeness of the wave vectors $k_s$ and $q/2$ appearing in the ansatz~\eqref{eq:ansatz} for the wave function].

It is also interesting to understand how the phase diagram discussed above depends on the density and on the interaction strengths of the BEC.
For a fixed ratio $g_2/g_1$, a larger density $\bar{n}$ tends to enhance the effects of the interactions. In particular, they favor the polarized
Bloch wave phase with respect to the unpolarized one, thus increasing the critical lattice strength $s_\mathrm{cr}$ for any given $\Omega$.
The same also happens for the critical Raman coupling $\Omega_\mathrm{tr}$ at fixed $s$, which leads to the enlargement of the region of
the mixed phase, in agreement with Ref.~\cite{Chen2016}. Concerning the role of the coupling constants, an increase of the ratio $g_2/g_1$
for a given density favors the configurations with vanishing magnetic polarization, i.e., the mixed and the unpolarized Bloch wave phases. This
yields a smaller value of $s_\mathrm{cr}$ and a larger critical Raman coupling $\Omega_\mathrm{tr}$. Notice that the latter result agrees with
the prediction of Eq.~\eqref{eq:Omega_tr_0} for the $s = 0$ value of $\Omega_\mathrm{tr}$. We finally point that, unlike the mixed phase, the
unpolarized Bloch wave phase can also appear in the presence of ferromagnetic spin-dependent interactions, i.e., when $g_2 < 0$.

\subsection{Role of magnetic fluctuations}
\label{subsec:mag_fluct}
\begin{figure}
\includegraphics{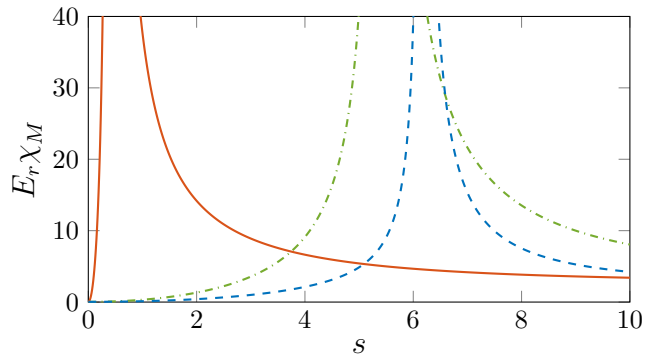}
\caption{Magnetic susceptibility $\chi^{}_M$ as a function of the dimensionless lattice strength $s$. The parameters are the same
as in Fig.~\ref{fig:ks_sz_contr_bulk}, i.e., $\Omega/E_r = 1.0$; $q = 1.5k_1^0$ (green dash-dotted line), $q = 2k_1^0$ (red solid line),
and $q = 3k_1^0$ (blue dashed line); $\bar{n} g_1/ E_r = 0.4$; and $g_2/g_1 = 0.0012$.}
\label{fig:chi_mag_bulk}
\end{figure}

A further important signature of the magnetic phase transition occurring in our system is given by the behavior of the magnetic susceptibility
$\chi^{}_M$. This can be evaluated in a way similar to the one described in Sec.~\ref{subsec:roton_dens_resp} for the compressibility:
One adds a small perturbation $V_\lambda = -\lambda \sigma_z$ to the single-particle Hamiltonian~\eqref{eq:soc_ham} and calculates the
expectation value $\langle\sigma_z\rangle_\lambda$ of the longitudinal spin operator on the perturbed ground state. Then the magnetic
susceptibility is given by $\chi^{}_M \equiv \dif\langle\sigma_z\rangle_\lambda/\dif\lambda|_{\lambda=0}$. The results show that $\chi^{}_M$
exhibits a divergent behavior at the transition between the polarized and the unpolarized Bloch wave phases (see Fig.~\ref{fig:chi_mag_bulk}),
revealing the occurrence of strong magnetic fluctuations. 

Notice that, if one considers values of $q$ too close to the roton momentum, such as $q=2k_1^0$ (red solid line in Fig.~\ref{fig:chi_mag_bulk}),
the value of the magnetic susceptibility grows very rapidly as one ramps up the lattice intensity starting from $s = 0$. Furthermore, $\chi^{}_M$
remains large for a wide range of values of $s$ much bigger than $s_\mathrm{cr}$. As a consequence, the curve $\langle\sigma_z\rangle$ vs $s$ of
Fig.~\ref{fig:ks_sz_contr_bulk}(b) and the phase diagram of Fig.~\ref{fig:phase_diag_Omega_s}(b) are significantly affected by the presence of
a nonvanishing effective magnetic field $\delta$ or by an even tiny difference between the coupling parameters $g_{\uparrow\uparrow}$ and
$g_{\downarrow\downarrow}$, as in the case of $^{87}$Rb. Figure~\ref{fig:chi_mag_bulk} also shows that the increase of the magnetic susceptibility
is notably slower in the $q=1.5k_1^0$ (green dash-dotted line) and especially in the $q=3k_1^0$ (blue dashed line) cases. Hence, the phase diagram
is much more stable if one takes the momentum transfer $q$ significantly different from the roton wave vector. We finally mention that the enhancement
of $\chi^{}_M$ can be further reduced by working at larger values of the Raman coupling $\Omega$, as we do in the following section.

\subsection{Magnetic phase transition in harmonic traps}
\label{subsec:harm_trap}
The results discussed so far were based on a calculation in the thermodynamic limit. However, it is important to understand how they are modified
if one considers a finite system in the presence of a trapping potential of harmonic type. For this purpose we have solved numerically the
three-dimensional Gross-Pitaevskii equation for a gas of $N = 1.8 \times 10^5$ $^{87}$Rb atoms, in the presence of an elongated trap with frequencies
$(\omega_x,\omega_y,\omega_z) = 2\pi \times (50,50,140)$ Hz. The scattering lengths are $a_{\uparrow\uparrow} = 100.83 a_B$ and
$a_{\downarrow\downarrow} = a_{\uparrow\downarrow} = 100.37 a_B$, $a_B$ being the Bohr radius. The value of the momentum transfer due to the Raman
laser is set to $k_0 = 5.52$ $\mu$m$^{-1}$. All the above parameters correspond to the conditions of the experiment of Ref.~\cite{Lin2011}.
For this calculation we have taken the Raman coupling $\Omega/E_r = 3.0$ and the momentum transfer $q = 3k_1^0 = 1.98 k_0$. This choice corresponds
to working close to the minimum of the critical lattice intensity $s_\mathrm{cr}$ appearing in Fig.~\ref{fig:phase_diag_Omega_s}(c); furthermore,
following the discussion in Sec.~\ref{subsec:mag_fluct}, one can expect that magnetic fluctuations will be less important in this regime of the
parameters.

\begin{figure}
\includegraphics{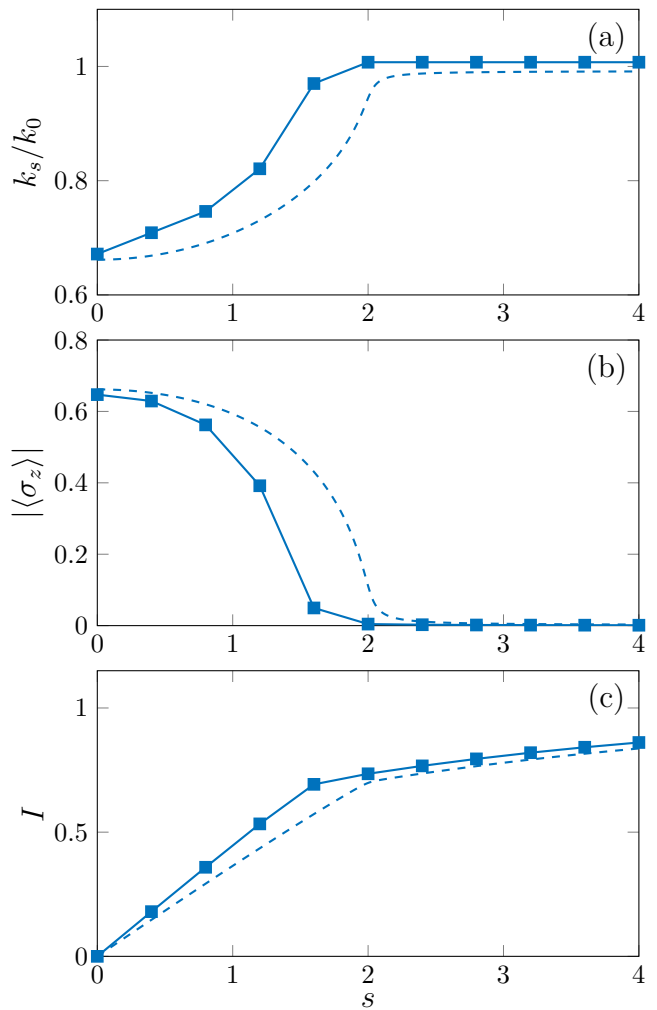}
\caption{(a) Condensation quasimomentum, (b) magnetic polarization, and (c) contrast of the fringes as a function of the dimensionless
lattice strength $s$ for $\Omega / E_r = 3.0$ and $q = 3k_1^0 = 1.98 k_0$. The dashed line corresponds to the case of an infinite system with
$\bar{n} g_1 / E_r = 0.4$, $g_2/g_1 = 0.0012$, and $g_3/g_2 = 2.0$. The squares are the results for a three-dimensional trapped gas in the
conditions described in the text, which yield a density at the center of the trap equal to the average density $\bar{n}$ of the infinite system.}
\label{fig:ks_sz_contr_trap}
\end{figure}

In Fig.~\ref{fig:ks_sz_contr_trap} we report our predictions (blue squares) for the quasimomentum [Fig.~\ref{fig:ks_sz_contr_trap}(a)], the magnetic
polarization [Fig.~\ref{fig:ks_sz_contr_trap}(b)], and the contrast of the fringes in the integrated density $n_1(x) = \int \dif y \, \dif z \, n(\vec{r})$
close to the center of the sample [Fig.~\ref{fig:ks_sz_contr_trap}(c)]. Notice that, since $g_3 > 0$ for the above values of the scattering lengths,
the quasimomentum and the magnetic polarization must actually be negative; in the figure we only plot their magnitudes. Our results clearly show that
the emergence of the magnetic phase transition can be detected also in trapped configurations. Furthermore, we have checked that, at variance with
the choice of small $\Omega$ and of $q$ too close to the roton momentum, the results of the simulations are not significantly perturbed by the inclusion
of a small but finite effective magnetic field $\delta$. 

Also in Fig.~\ref{fig:ks_sz_contr_trap} we also report the results (blue dashed lines) for an infinite system whose average density $\bar{n}$
coincides with the density at the center of the trapped BEC described above. By comparing the two sets of data, one immediately sees that the
results for the two systems are slightly different from the quantitative point of view. Indeed, as a consequence of the spatially varying density,
the values of the magnetic polarization in the trapped gas are smaller than in the infinite system and they decrease more rapidly with increasing
$s$. However, the qualitative behavior is very similar to the one in the absence of the trap.

\section{Conclusion}
\label{sec:conclusion}
The static response to a weak density perturbation of a spin-orbit-coupled Bose-Einstein condensate
in the plane-wave phase exhibits an important dependence on the wave vector of the external potential.
If the latter is close to the momentum at which the roton minimum occurs, the response acquires a very
large value close to the transition to the striped phase. This phenomenon is related to the vanishing of
the energy of the roton minimum and reveals the presence of important highly nonlinear effects. By
studying the ground state of the system in the presence of a one-dimensional optical lattice, we have shown
that such effects are strongly connected with the occurrence of a second-order magnetic phase transition
to an unpolarized Bloch wave phase. In this state the contrast of the density modulations can achieve much
larger values than in the striped phase appearing at low Raman couplings in the absence of the lattice. The
critical lattice intensity needed to achieve the magnetic phase transition can assume very tiny values if
the lattice wave vector is close to the roton momentum; in the opposite case, the transition can take place
at larger lattice strengths. The phase transition is accompanied by a divergent behavior of the magnetic
susceptibility. The stability of the system against magnetic fluctuations can be enhanced by choosing the
momentum transfer due to the optical lattice far enough from the roton momentum and considering large values
of the Raman coupling. Our predictions for infinite systems have been confirmed also by Gross-Pitaevskii
calculations in a three-dimensional trap with realistic parameters, which opens the possibility of exploring
these nonlinear phenomena in current experiments.

\begin{acknowledgments}
Useful correspondence and discussions with Peter Engels, Amin Khamehchi, and Yun Li are acknowledged.
G.I.M. was partially supported by the PRIN Grant No.~2010LLKJBX ``Collective quantum phenomena:
From strongly correlated systems to quantum simulators'' and by INFN through the project ``QUANTUM.''
The research leading to these results received funding from the European Research Council under
European Community's Seventh Framework Programme (FR7/2007-2013 Grant Agreement No. 341197).
T.O., C.Q., and S.S. were supported by ERC through the QGBE grant, by the QUIC grant of the Horizon2020
FET program, and by Provincia Autonoma di Trento. T.O. was also supported by the EU-FET Proactive grant
AQuS, Project No. 640800.
\end{acknowledgments}

\appendix*
\section{Band structure of the ideal Bose gas}
\label{sec:band_str}
As discussed in Sec.~\ref{subsec:mb_ground_state}, in the absence of the periodic potential~\eqref{eq:periodic_pot} and for $\Omega >
\Omega^0_\mathrm{tr}$, the minima of the energy~\eqref{eq:soc_energy} of a spin-orbit-coupled BEC occur at two momenta $\pm k_1$,
with $k_1$ given by Eq.~\eqref{eq:k1}. In Sec.~\ref{subsec:unmixed_regime} it was pointed out that, by increasing the intensity $s$ of
the periodic potential, one forces the system to adapt its minima to become closer to the wave vector $q/2$ of the external periodic
potential. Eventually, when the separation between the two minima equals $q$, i.e., when the two minima coincide with the edges of the
Brillouin zone, they correspond to the same physical state. As a consequence, the magnetic polarization disappears.

\begin{figure}
\vspace*{5mm}
\includegraphics{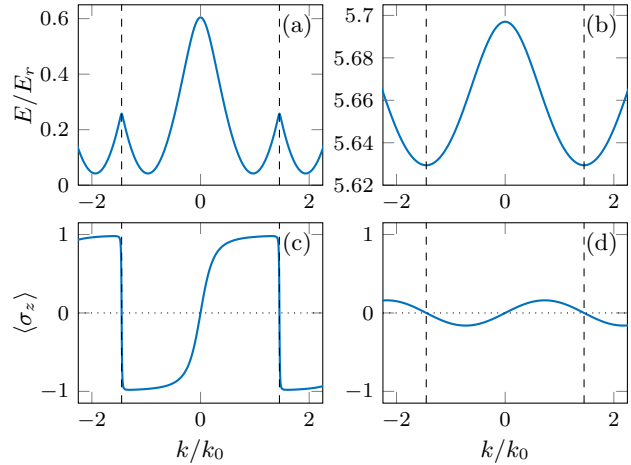}
\caption{(a) and (b) Lowest-lying energy band of the single-particle dispersion of a spin-orbit-coupled BEC in a one-dimensional optical
lattice and (c) and (d) the corresponding magnetic polarization as a function of the quasimomentum $k$, for $\Omega/E_r = 1.0$
and $q = 3k_1^0 = 2.9 k_0$. The left and right panels show the results for $s = 0.1$ and $s=10.0$, respectively. The region between
the vertical dashed lines corresponds to the first Brillouin zone.}
\label{fig:band_structure}
\end{figure}

The above mechanism is simply understood in the ideal Bose gas model described by the spin-orbit Hamiltonian~\eqref{eq:soc_ham} with
the addition of the lattice potential~\eqref{eq:periodic_pot}~\cite{Zhang2013}. In Fig.~\ref{fig:band_structure} we show the lowest-lying
band of the single-particle dispersion for $\Omega/E_r = 1.0$, $q = 3k_1^0 = 2.9 k_0$, and two different values of $s$. If $s$ is small,
the distance of the two minima at $k=\pm k_s$ significantly differs from $q$ and they carry opposite magnetic polarizations given by
$\sim \pm k_s/k_0$, as shown in Figs.~\ref{fig:band_structure}(a) and \ref{fig:band_structure}(c), respectively. By increasing the value
of $s$ the two minima eventually approach the edge of the Brillouin zone and the polarization disappears [see Figs.~\ref{fig:band_structure}(b)
and \ref{fig:band_structure}(d)]. Although these examples refer to a situation with $q > 2k_1^0$, a similar behavior can take place in the
opposite case of $q < 2k_1^0$, i.e., when the value of $k_1^0$ does not lie in the first Brillouin zone (see the discussion in
Sec.~\ref{subsec:unmixed_regime}).

Finally, it is worth noting that, if one chooses $q = 2 k_1^0$, the single-particle dispersion minima are located at the edge of the Brillouin
zone for any nonvanishing value of $s$. Thus, the ground state in the ideal Bose gas model is unpolarized as soon as $s \neq 0$.

\end{document}